\documentclass[twocolumn,showpacs,preprintnumbers,amsmath,amssymb,superscriptaddress]{revtex4-1}
\usepackage{graphicx}
\usepackage{dcolumn}
\usepackage{bm}
\usepackage{amsmath}

\begin{document}

\title{Equivalent parameters for series thermoelectrics}

\author{Y. Apertet}\email{yann.apertet@gmail.com}
\affiliation{Institut d'Electronique Fondamentale, Universit\'e Paris-Sud, CNRS, UMR 8622, F-91405 Orsay, France}
\author{H. Ouerdane}
\affiliation{Russian Quantum Center, 100 Novaya Street, Skolkovo, Moscow region 143025, Russia}
\affiliation{Laboratoire Interdisciplinaire des Energies de Demain (LIED), UMR 8236 Universit\'e Paris Diderot, CNRS, 5 Rue Thomas Mann, 75013 Paris, France}
\author{C. Goupil}
\affiliation{Laboratoire Interdisciplinaire des Energies de Demain (LIED), UMR 8236 Universit\'e Paris Diderot, CNRS, 5 Rue Thomas Mann, 75013 Paris, France}
\author{Ph. Lecoeur}
\affiliation{Institut d'Electronique Fondamentale, Universit\'e Paris-Sud, CNRS, UMR 8622, F-91405 Orsay, France}

\date{\today}

\begin{abstract}
We study the physical processes at work at the interface of two thermoelectric generators (TEGs) thermally and electrically connected in series. We show and explain how these processes impact on the system's performance: the derivation of the equivalent electrical series resistance yields a term whose physical meaning is thoroughly discussed. We demonstrate that this term must exist as a consequence of thermal continuity at the interface, since it is related to the variation of the junction temperature between the two TEGs associated in series as the electrical current varies. We then derive an expression for the equivalent series figure of merit. Finally we highlight the strong thermal/electrical symmetry between the parallel and series configurations and we compare our derivation with recent published results for the parallel configuration.
\end{abstract}

\maketitle

\section{Introduction}

Optimization of thermoelectric systems for energy conversion applications involves various strategies for improvement of the materials' properties to enhance their so-called figure of merit, device design and working conditions \cite{Rowe2006,Apertet2012b}. Besides carrier-doping, nanostructuring techniques and use of multi-phase structures (see, e.g., recent works in Refs.~\cite{Parker2012,Ke2009,Gelbstein2012} and further references therein) to enhance electrical conductivity on the one hand and reduce lattice conductivity on the other hand, the segmentation of thermoelectric legs in a module is one of the strategies used to yield the best possible performance out of the materials. A segmented leg is obtained by stacking different thermoelectric materials in such a way that each part of the leg is optimized for the temperature that it experiences within the structure. The basic principle of this technique lies in the temperature-dependence of the materials' properties: a given material presents interesting conversion capabilities only within a limited temperature range. So, when the figure of merit of a given material collapses at some point of the leg because of the  variation of temperature along the leg, this material is replaced at this point by another one, better suited to the local temperature. The benefits on the global performance of the device is obvious when the temperature difference imposed on the TEG is important \cite{ElGenk2003}. A high performance leg is composed of different segments thermally and electrically in series.

The principle of segmentation was patented in 1962 \cite{Fredrick1962}. However this segmentation strategy is not always the best solution to achieve high power efficiency as it was pointed out many times and even before the publication of the patent. Thus, along the years, many criteria were selected and used to see if the global efficiency could sufficiently increase by segmentation. If some studies focused on local entropy production \cite{Clingman1961} and differential output power \cite{Schilz1998}, others developed more global approaches \cite{Harman1958, Heikes1961}. Fairly recently Snyder and Ursell proposed the so-called compatibility approach giving a rule of thumb based on a compatibility factor associated with each material \cite{Ursell2002, Snyder2003}.

Our purpose here is to highlight the underlying physical mechanisms permitting or not compatibility between different thermoelectric materials. We focus on the simple, yet illustrative, case of a leg composed of two thermoelectric segments. The key issue that we address is the determination of the temperature at the junction, $T_{\rm m}$, and its variation as the system operates. This was hardly ever considered even though El-Genk and Saber noticed that the electrical current should have an impact on this temperature ~\cite{ElGenk2003}. In the present work, we consider constant parameters to remain on an analytical level and hence put forth the essential physical processes at work in the thermoelectric transport at the interface of the two subsystems; but note that our analysis may extend to cases with temperature-dependent materials's parameters.

Our paper is organized as follows. In Sec.~\ref{sec:equivalent}, we present our model of a two-segment thermoelectric generator; we focus especially on the equivalent series parameters and figure of merit, the physical meaning of which is discussed. Then, in Sec.~\ref{sec:compatibility}, we compare our derivations to the so-called compatibility approach of the segmented generators, demonstrating that the latter works only under restrictive conditions. In Sec.~\ref{sec:comparison}, we relate the system treated in this article to a device composed of two TEGs thermally and electrically in parallel: we highlight the thermal/electrical symmetry between the two configurations, in agreement with Mahan's recent results \cite{Mahan2013}. We end this paper with concluding remarks and an Appendix where we show that the neglect of Joule heating and transferred power to the load in our model, is of no importance in the calculation of the junction temperature. 

\section{\label{sec:equivalent}Equivalent model of two thermoelectric generators in series}

\subsection{Definitions}

\begin{figure}
	\centering
		\includegraphics[width=0.48\textwidth]{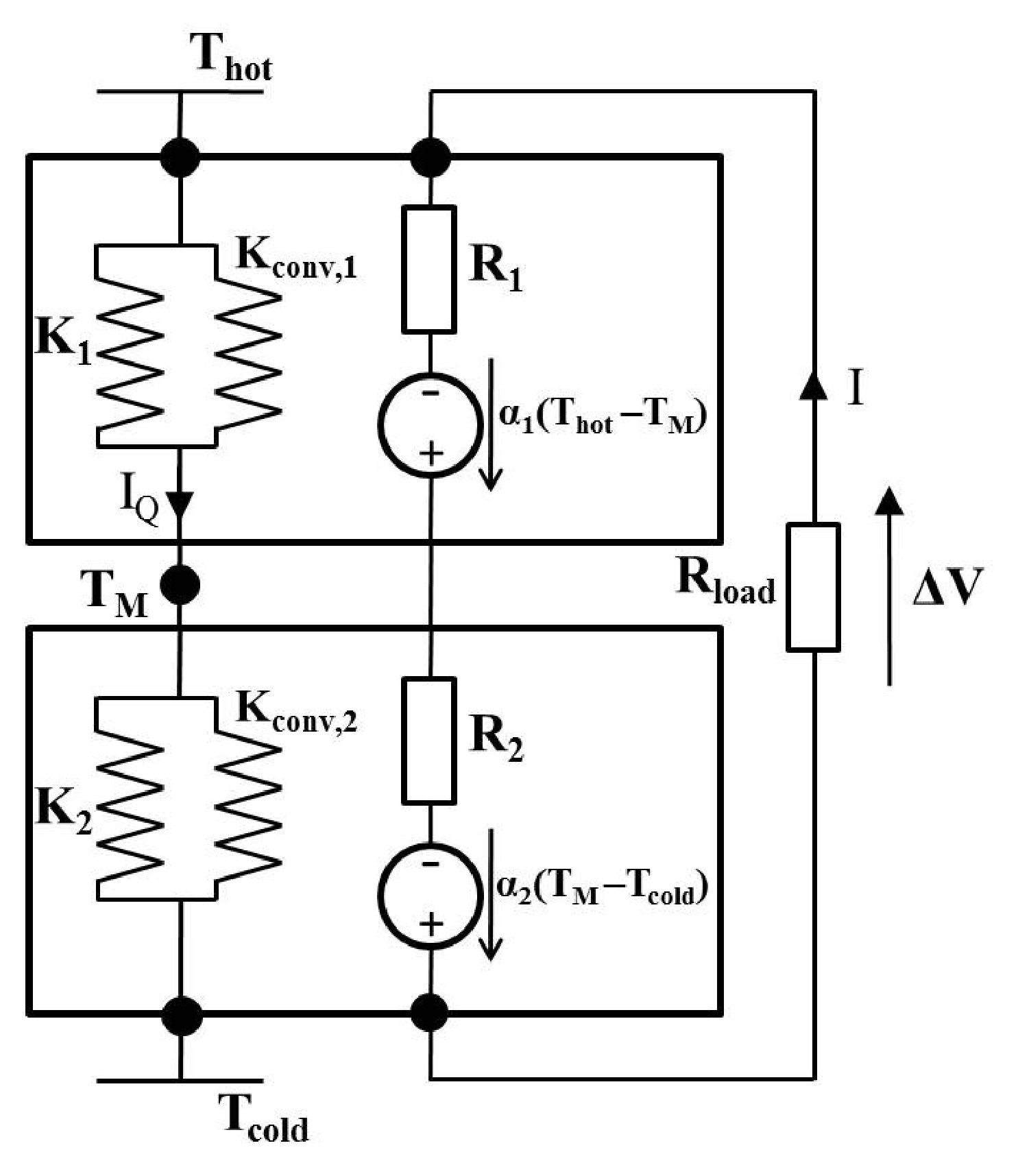}
	\caption{Schematic representation of a segmented thermoelectric generator composed of two parts thermally and electrically connected in series.}
	\label{fig:figure1}
\end{figure}

We consider two thermoelectric generators labeled 1 and 2, electrically and thermally connected in series as depicted in Fig.~\ref{fig:figure1}. The electrical current through the resistive load $R_{\rm load}$, is $I$. Each generator is characterized by an internal electrical resistance $R_i$, a Seebeck coefficient $\alpha_i$, and a thermal conductance under open-circuit condition $K_i$ ($i = 1,2$). The voltages and temperature differences across each module are denoted $\Delta V_i$ and $\Delta T_i$. The electrical currents and average thermal fluxes through each TEG, $I_i$ and $I_{Q_i}$, are derived from the phenomenological force-flux formalism \cite{Callen1948}:

\begin{equation}\label{frcflx}
\left(
\begin{array}{c}
I_i\\
I_{{Q}_{i}}\\
\end{array}
\right)
= \frac{1}{R_i}
\left(
\begin{array}{cc}
1~ & ~\alpha_i\\
\alpha_i T~ & ~\alpha_i^2 T + R_i K_{i}\\
\end{array}
\right)
\left(
\begin{array}{c}
\Delta V_i\\
\Delta T_i\\
\end{array}
\right)
\end{equation}
\noindent where $T$ is the average temperature of the whole system. The figure of merit of each TEG is $Z_iT = \alpha_i^2 T / (R_i K_i)$.

The mean heat flux in TEG $i$ reads:

\begin{equation}
I_{{Q}_{i}} = \alpha_i T I_{i} + K_i \Delta T_i
\end{equation}

\noindent where the Peltier term $\alpha_i T I_{i}$ is a quantitative measure of the heat transferred by thermoelectric convection process \cite{Thomson1856, Apertet2012c}; so as depicted in Fig.~\ref{fig:figure1}, the heat conveyed by the electrical current is represented by the additional thermal conductances $K_{\rm conv,1}$ and $K_{\rm conv,2}$. These conductances depend on electrical load through the electrical current $I$ and may be expressed as \cite{Apertet2012b}:

\begin{equation}
K_{\rm conv,i} = \frac{\alpha_i T I_i}{\Delta T_i}
\end{equation}

\noindent The constitutive equations of this model are derived assuming that the heat flux is constant along each TEG and equal to its average value. This assumption amounts to neglecting both the Joule heating and the electrical power produced, i.e., considering a TEG with small efficiency (the validity of these assumptions is discussed in the Appendix). Here, we further make the approximation that the mean temperature $T$ can be considered as constant through the system as it is high compared to the temperature difference between the heat reservoirs. All these assumptions are consistent with the linear approximation used to describe the behavior of the segments.

\subsection{Temperature at the junction}

Continuity of the heat flux at the junction yields:

\begin{equation}
K_1 (T_{\rm hot} - T_{\rm m}) + \alpha_1 T I = K_2 (T_{\rm m} - T_{\rm cold}) + \alpha_2 T I
\end{equation}

\noindent from which we derive the expression for the temperature at the junction between the two segments, $T_{\rm m}$:

\begin{equation} \label{Tm}
T_{\rm m} = \frac{1}{K_1 + K_2} \left[(\alpha_1 - \alpha_2) T I + K_1 T_{\rm hot} + K_2 T_{\rm cold}\right]
\end{equation}

\noindent A more detailed description of the continuity of the heat flux is given in the Appendix.

Let us now focus on the electrical part of the generator. The voltage across the whole module is the sum of the voltages across each segment:

\begin{equation}
\Delta V = \alpha_1 (T_{\rm hot} - T_{\rm m}) + \alpha_2 (T_{\rm m} - T_{\rm cold}) - (R_1 + R_2)I
\end{equation}

\noindent Substitution of $T_{\rm m}$ by its full expression (\ref{Tm}) yields

\begin{eqnarray}
\nonumber
\Delta V & = & \frac{K_2 \alpha_1 + K_1 \alpha_2}{K_1 + K_2} (T_{\rm hot} - T_{\rm cold})\\
& - &\left[\frac{(\alpha_1 - \alpha_2)^2 T}{K_1 + K_2}+ R_1 + R_2\right] I
\end{eqnarray}

\noindent which assumes the same form as that expected for a Th\'evenin generator:

\begin{equation}
\Delta V = \alpha_{\rm eq} (T_{\rm hot} - T_{\rm cold}) - R_{\rm eq} I
\end{equation}

\noindent where $\alpha_{\rm eq}$ is the equivalent series Seebeck coefficient and $R_{\rm eq}$ is the equivalent series electrical resistance:

\begin{equation}
\label{alphaeq}
\alpha_{\rm eq} = \frac{K_2 \alpha_1 + K_1 \alpha_2}{K_1 + K_2}
\end{equation}

\noindent and

\begin{equation}
\label{Req}
R_{\rm eq} = R_1 + R_2 + R_{\rm relax}
\end{equation}

\noindent In open circuit condition we find that the equivalent thermal conductance $K_{\rm eq}$ for the whole system is given by
\begin{equation}
\label{Keq}
K_{\rm eq} = \frac{K_2 K_1}{K_1 + K_2},
\end{equation}

\noindent which is a standard form for an equivalent series conductance; however, the finite third term in Eq.~(\ref{Req}):

\begin{equation}
\label{Rrelax}
R_{\rm relax} = \frac{(\alpha_1 - \alpha_2)^2 T}{K_1 + K_2}
\end{equation}

\noindent is rather unexpected but, as it turns out, much insight into the physics of device operation may be derived from it.  

\subsection{On the meaning of $R_{\rm relax}$}

The resistance $R_{\rm relax}$ is proportional to the difference between the Seebeck coefficients $\alpha_1$ and $\alpha_2$, which clearly shows that in presence of materials with significantly different thermopowers, the standard form for the equivalent series resistance does not apply; further, note that $R_{\rm relax}$ is by no means related to an electrical contact resistance at the interface between the two segments, which we neglect in our model. Its appearance in Eq. (\ref{Req}) is directly related to the relaxation of the temperature at the junction $T_{\rm m}$, which depends on the value of the electrical current $I$. Indeed, to ensure thermal continuity at the interface, when the convective part of the thermal current varies, the conductive part must change accordingly to account for the modification of the temperature at the junction.

Thus when for example $\alpha_1 < \alpha_2$, as the electrical current is constant along the whole system, the conductive part of the thermal flux ($\alpha_i T I$) is larger in the second segment than it is in the first one. Thus, to satisfy the condition of continuity of the \emph{total} thermal flux, the conductive part ($K_i \Delta T_i$) must be larger in the first segment than it is in the second one. This thermal balance can only be obtained by a relaxation of the temperature $T_{\rm m}$. Since this difference between the convective thermal fluxes in each segment increases when the electrical current increases, $T_{\rm m}$ varies with the electrical current $I$, as shown in Eq.~(\ref{Tm}). Such a temperature relaxation at the junction always yields an increase of the temperature difference across the segment with the lower Seebeck coefficient (i.e., the one with less capability to produce power); conversely, a decrease of the temperature difference across the segment with the higher Seebeck coefficient is observed: the \emph{global} electromotive force developed by the whole system is therefore reduced. This detrimental effect for energy conversion performances is reflected by the existence of the additional term $R_{\rm relax}$.

To further discuss the consequences of the presence of $R_{\rm relax}$ in Eq.~(\ref{Req}), we propose a numerical example. The parameters for TEG~2 are set to the following values: $Z_2T = 1$, $K_2 = 2.5$~mW/K, $R_2 = 4.8~$m$\Omega$, and $\alpha_2 = 200 ~\mu$V/K. These parameters are consistent with the properties at room temperature of a particular bismuth telluride compound, (Bi$_{0.25}$Sb$_{0.75}$)$_2$Te$_3$, one of the most efficient thermoelectric materials \cite{Yamashita2003}. For TEG~1, we make the assumption that the material used has a fixed figure of merit $Z_1T = 2$, which is optimistic but remains feasible \cite{Vining2009}. $K_1$ is also fixed, with $K_1 = K_2$. The thermopower $\alpha_1$ is variable and so is $R_1$ since it is matched to the value of $Z_1T$. The reservoirs temperatures are $T_{\rm hot} = 305$ K and $T_{\rm cold} = 295$~K, and the mean temperature is $T = 300$~K. Note that all the numerical results and corresponding plots presented in this paper are obtained with these parameters.

\begin{figure}
	\centering
		\includegraphics[width=0.48\textwidth]{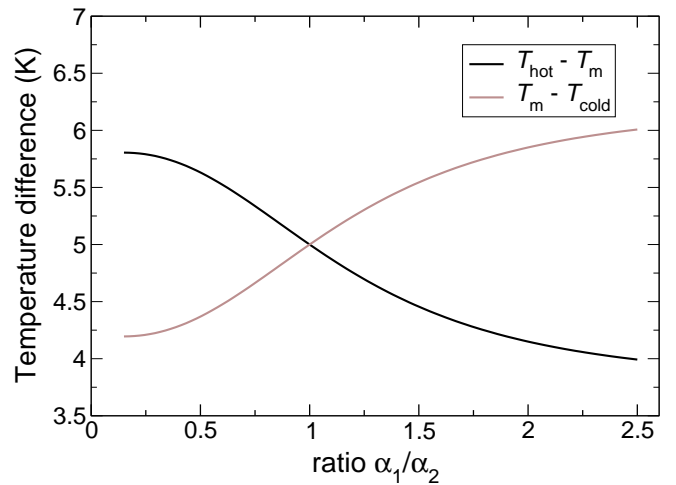}
	\caption{Temperature difference at the edges of each segment when the global system is at maximum efficiency as a function of $\alpha_1$ (normalized by $\alpha_2$).}
	\label{fig:figure2}
\end{figure}

Figure~\ref{fig:figure2} displays the dependence on $\alpha_1$ of the temperature difference experienced by each module when the whole system works at maximum efficiency. The thermal conductances of each TEG are identical; this specification allows to obtain an equal partition of the temperature difference when there is no electrical current inside the structure. When the Seebeck coefficients are identical there is, as expected, no relaxation of $T_{\rm m}$, but when $\alpha_1$ differs from $\alpha_2$, we notice that one segment experiences a larger temperature difference. The favored side is always that with the smaller Seebeck coefficient.

\subsection{Equivalent series figure of merit}

Now that we have derived the equivalent parameters we may express the equivalent series figure of merit $Z_{\rm eq}$ as follows:

\begin{equation}
Z_{\rm eq} = \frac{\alpha_{\rm eq}^2}{R_{\rm eq} K_{\rm eq}}
\end{equation}

\noindent which, using Eqs.~(\ref{alphaeq}),(\ref{Req}) and (\ref{Keq}), may be rewritten as the product of two terms:

\begin{equation}
\label{Zeq}
Z_{\rm eq} = Y Z_{\rm series}
\end{equation}

\noindent with

\begin{equation}
Z_{\rm series} = \left(\frac{\displaystyle K_2 \alpha_1 + K_1 \alpha_2}{\displaystyle K_1 + K_2}\right)^2\left[\frac{\displaystyle K_1 K_2}{\displaystyle K_1 + K_2}(R_1 + R_2)\right]^{-1}
\end{equation}

\noindent and

\begin{equation}
Y = \left[1 + \frac{\displaystyle \left(\alpha_1 - \alpha_2\right)^2 T}{\displaystyle (R_1 + R_2)(K_1 + K_2)}\right]^{-1}
\end{equation}

The term $Z_{\rm series}$ should be seen as an \emph{expected} term for which the equivalent series electrical resistance is given by the standard sum of $R_1$ and $R_2$. The factor $Y$, on the contrary, is purely related to the relaxation of the temperature at the junction between the two TEGs. This factor reflects the fact that this relaxation always leads to a decrease of the performance; it is indeed smaller than $1$ except for $\alpha_1 = \alpha_2$. It is also interesting to note that the thermal conductances and electrical resistances also appear in the expression of $Y$ and that these quantities must be small for $Y$ to have a significant impact on $Z_{\rm eq}$.

\begin{figure}
	\centering
		\includegraphics[width=0.48\textwidth]{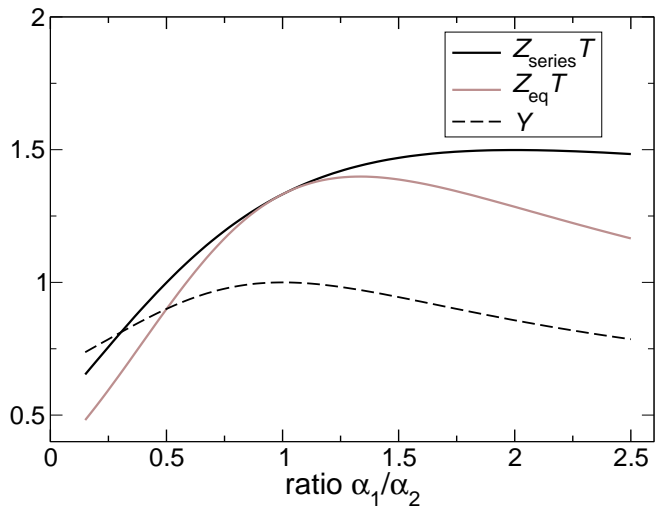}
	\caption{$Z_{\rm eq} T$, $Z_{\rm series} T$ and $Y$ plotted against the ratio $\alpha_1 / \alpha_2$.}
	\label{fig:figure3}
\end{figure}

Figure~\ref{fig:figure3} obtained for the same numerical parameters as above, illustrates the fact that in order to optimize $Z_{\rm eq}$, one has to make a compromise between the mere association of the segments, represented by $Z_{\rm series}$, and the effect resulting from the mismatch between the Seebeck coefficients, represented by $Y$.

To end this section on the equivalent series figure of merit, we want to stress that Bergman and Levy's theorem is satisfied: $Z_{\rm eq}T$, given by Eq.~(\ref{Zeq}), is always smaller than that of the segment with the highest figure of merit \cite{Bergman1991}.

\section{\label{sec:compatibility}Discussion of the thermoelectric compatibility approach}
The optimization of segmented TEGs is inseparable of the concept of thermoelectric compatibility developed by Snyder and Ursell \cite{Snyder2003}. We show here that our approach is, to some extent, different from that of thermoelectric compatibility, and we explain why we observe a discrepancy between the results given by both methods.

To evaluate the difference between the two approaches, we first remind the definitions of the physical quantities pertaining to the thermoelectric compatibility. The relative current $u_i$ for TEG $i$ is defined as:

\begin{equation}
u_i = \frac{J}{\kappa_i \nabla T_i}
\end{equation}

\noindent where $J$ is the electrical current density and $\kappa_i$ the thermal conductivity under open-circuit condition. The macroscopic expression for the relative current is:

\begin{equation}
u_i = \frac{I}{K_i \Delta T_i}
\end{equation}

The compatibility factor $s$, namely the optimal value for the relative current $u$ is expressed as:
\begin{equation} \label{compatibilityfactor}
s_i = \frac{\sqrt{1+ Z_i T} - 1}{\alpha_i T}
\end{equation}

\noindent Two materials are considered compatible only if their compatibility factors are close (within a factor of 2) \cite{Ursell2002}. This condition can be viewed as a necessity to have optimal values of the electrical currents in each segment, which are close enough: in this way, as the current is the same through the device, each segment works optimally at the same time \cite{Vikhor2009}.

\begin{figure}
	\centering
		\includegraphics[width=0.48\textwidth]{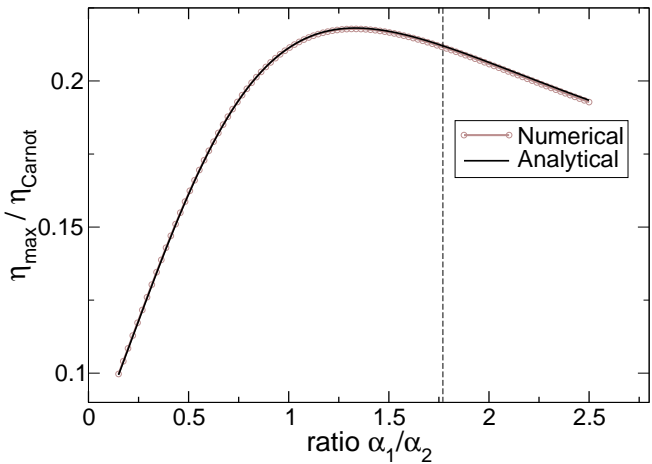}
	\caption{Maximum efficiency (scaled the Carnot efficiency) as a function of $\alpha_1$ evaluated both numerically and from Eq.~(\ref{effmax}).}
	\label{fig:figure4}
\end{figure}

Our method is based on the optimization of the equivalent series figure of merit $Z_{\rm eq}T$ defined in Eq.~(\ref{Zeq}) and shown against the thermopowers ratio $\alpha_1/\alpha_2$, in Fig.~\ref{fig:figure3} for the particular case considered in this paper. To check that $Z_{\rm eq}T$ is a meaningful quantity we evaluate the maximum efficiency using the standard analytical form \cite{Ioffe1957}:  

\begin{equation} \label{effmax}
\eta_{\rm max} = \eta_{\rm Carnot} \frac{\sqrt{1 + Z_{\rm eq}T} - 1}{\sqrt{1 + Z_{\rm eq}T} + \frac{T_{\rm cold}}{T_{\rm hot}}}
\end{equation}

\noindent The result is shown in Fig.~(\ref{fig:figure4}). For comparison we also plot the maximum efficiency calculated with a numerical simulation using the complete description of the device given in Eq.~(\ref{frcflx}). The two curves are very close so we may conclude that $Z_{\rm eq}T$ is a proper parameter to estimate the efficiency capability of the whole device.  

With the case considered in this paper, the value of $\alpha_1$ that permits satisfaction of the condition $s_1 = s_2$, is given by $\alpha_1 = \alpha_2 \frac{\sqrt{1 + Z_1T} - 1}{\sqrt{1 + Z_2T} - 1} = 1.77~\alpha_2$ [vertical dotted-line in Fig.~(\ref{fig:figure4})] whereas with our equivalent model we find an optimal value of $\alpha_1 = 1.34~\alpha_2$, which, when compared to the numerical result, leads to a better optimization for the maximum efficiency.

Why the compatibility approach fails to accurately predict the optimal set of parameters in this particular case? To derive their expression of the compatibility factor $s$, it appears that Snyder and Ursell assumed that heat transport by conduction remains constant along the device~\cite{Snyder2003}:\\
``\emph{Since all segments in a thermoelectric element are electrically and thermally in series, the same current I and similar conduction heat $A \kappa \nabla T$ flow through each segment}''
\\
This assumption is more explicitly stated in Ref.~\cite{Vikhor2009}. However, as soon as the figure of merit varies from one segment to the next one, the partition of heat flux between the convective and the conductive contributions changes along the device. As derived in Ref.~\cite{Apertet2012c}, $ZT$ may indeed be related to the ratio of these two contributions and it follows that, in order to guarantee the satisfaction of the condition of heat continuity (though, still neglecting the produced power and the Joule effect at the interface) the conductive part given by $K_i \Delta T_i$ is internally set to the appropriate value through the relaxation of $T_{\rm m}$. The compatibility approach thus looses accuracy when the figures of merit of each segment are both high (i.e. greater than 1) and relatively dissimilar, which is the case of the numerical example treated in this paper. The determination of the relative current at a local scale through numerical simulations remains a powerful tool for device optimization nonetheless, as it allows to deal with generators composed of more than two segments and with materials characterized by non constant parameters \cite{Vikhor2009, Zabrocki2010, Seifert2010}.

Our approach also offers the distinct advantage to provide a simple description of the whole thermoelectric system \emph{that remains valid for every working conditions} contrary to the compatibility approach. Indeed this latter only focuses on the maximum efficiency conditions and hence it fails to account for the other desirable load conditions such as maximum power output (see, for example the discussion in Ref.~\cite{Apertet2014}).

\section{\label{sec:comparison}Comparison between series and parallel configurations}

In a recent article \cite{Apertet2012a} we studied the association of two TEGs electrically and thermally connected in parallel: we demonstrated that if the Seebeck coefficients of the two segments composing the device are different, an internal electrical current may develop and yield an additional term for the equivalent parallel conductance under open circuit condition. This additional term, $K_{\rm conv}$, results from the convection process, i.e. heat that is conveyed by the electrical current inside the structure as defined by Thomson \cite{Thomson1856}.

The conductance $K_{\rm conv}$ reads

\begin{equation} \label{Kconv}
K_{\rm conv} = \frac{\left(\alpha_1 - \alpha_2\right)^2 T}{R_1 + R_2}
\end{equation}

\noindent At this stage, it is instructive to highlight the similar forms of Eq.~(\ref{Kconv}) and Eq.~(\ref{Rrelax}): both expressions are proportional to the square of the difference of the Seebeck coefficients, and one may switch from one to the other only by exchanging the nature of the physical property under consideration: one only needs to replace the electrical resistance by the thermal conductance and \emph{vice versa}. It is interesting to point out that the underlying symmetry between thermal and electrical transport in thermoelectric phenomena is reflected by the correspondence between these two equations. Indeed, we notice that the electrical transport impacts the thermal transport in the parallel configuration as much as the thermal transport impacts the response of the electrical circuit in the series configuration. 

We pursue the comparison between both cases: as for the series configuration, the equivalent parallel figure of merit may be expressed as the product of an \emph{expected} term and the factor $Y$ as defined in Section~\ref{sec:equivalent}:

\begin{equation}\label{zypara}
Z_{\rm eq}^{\parallel}T = Y \frac{G_1 + G_2}{K_1 + K_2} \left(\frac{\displaystyle G_1 \alpha_1 + G_2 \alpha_2}{\displaystyle G_1 + G_2}\right)^2~T
\end{equation}

\noindent which yields the following definition of the equivalent parallel thermopower and conductance:

\begin{equation}
\alpha_{\rm eq}^{\parallel}= \frac{G_1 \alpha_1 + G_2 \alpha_2}{G_1 + G_2} ~\mbox{and}~ G_{\rm eq}^{\parallel} = G_1 + G_2
\end{equation}

\noindent where for ease of notations, in the parallel case, we use the electrical conductance $G (= 1 / R)$ instead of the electrical resistance.

Lin-Chung and Reinecke \cite{LinChung1995}, and recently Mahan \cite{Mahan2013} found that the performance of thermoelectric systems both electrically and thermally connected in parallel, e.g., superlattice structures, must be characterized by two figures of merit: one derives from the averages of the thermopowers, and of the thermal conductivities; the other depends on the square of the difference of the thermopowers, and results in a global reduction of the composite system's efficiency. Our study not only shows that the performance of thermoelectric systems connected in series must also be characterized by an equivalent figure of merit that reads as the product of an \emph{expected} term and another term that negatively impacts of the global device performance, but that this latter assumes the same expression as that for the parallel configuration, as shown in Eq.~(\ref{zypara}). Further, our model permits further physical insight into the processes that yield the observed performance decrease.

\section{\label{sec:summary}Concluding remarks}

In the search for best possible thermoelectric properties, configuration/geometrical aspects at the materials' level are of importance as shown recently by Gelbstein who found that for a phase-separated alloy, series alignment yields optimal properties \cite{Gelbstein2012}. To analyze the physics of the series configuration at the system's level, we have presented a simple model of a segmented thermoelectric generator composed of only two segments. We demonstrated that when the Seebeck coefficients of each segment are different, the temperature at the junction changes as the electrical current varies. This effect is embodied in an additional term for the equivalent series electrical resistance, and reflects the associated decrease of performance. Knowledge of equivalent series parameters allowed the derivation of an equivalent series figure of merit, which accurately characterizes the efficiency capability of the whole device when the Joule heating contribution to the total heat flowing through the device remains negligibly small. Fortunately, this condition is satisfied for most of the segmented thermoelectric generators as discussed in Appendix A. Finally we have highlighted the symmetry that exists between the association of TEGs in parallel and the association in series. Extension of our study of the behavior of the junction temperature to small thermoelectric devices should account for high electrical current densities through the interface of the composite system and include an analysis of thermal boundary resistance, both of which reduce the efficiency of micro-devices \cite{Hao}.

\section*{Acknowledgments}

Y.A. acknowledges financial support from the Minist\`ere de l'Enseignement Sup\'erieur et de la Recherche.

\appendix

\section{\label{sec:annexeA} Approximate expression against exact expression}

\begin{figure}[h]
	\centering
		\includegraphics*[width=0.480\textwidth]{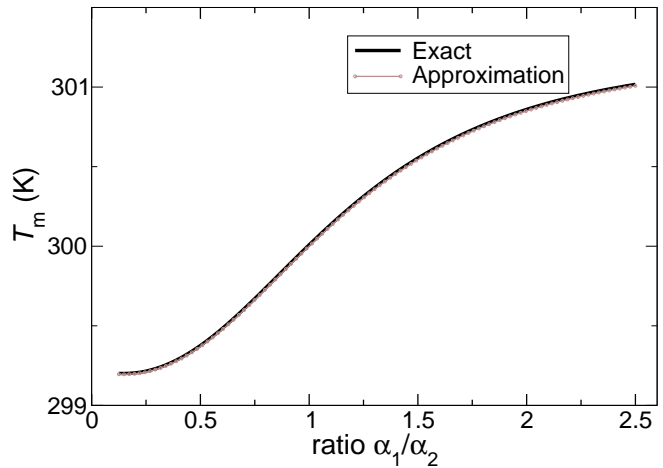}
	\caption{Comparison of the exact and approximated values of $T_{\rm m}$ for the maximum efficiency working condition as a function of $\alpha_1$.}
	\label{fig:figure5}
\end{figure}

In this article, for simplicity we used an approximate description of the TEGs, neglecting both Joule heating and power transferred to the load, in order to derive the expression of the  temperature $T_{\rm m}$. We show here that the approximations we made is justified. To do so, we compare the results obtained with Eq.~(\ref{Tm}) to those obtained from the following exact expression derived from the full condition of continuity of the heat flux at the junction:

\begin{equation}\label{Tmexact}
T_{\rm m} = \frac{K_1 T_{\rm hot} + K_2 T_{\rm cold} + \frac{(R_1 + R_2)}{2} I^2}{K_1 + K_2 + (\alpha_2 - \alpha_1) I}
\end{equation}

Retaining the same values of the parameters as those used for the numerical example treated in the main text, we computed the dependence of $T_{\rm m}$ on the thermopower $\alpha_1$ (scaled to $\alpha_2$), given by Eq.~(\ref{Tm}) and by Eq.~(\ref{Tmexact}) when the whole system works at maximum efficiency. The results are shown in figure~\ref{fig:figure5}. We notice that the approximation holds very well for the different values of $\alpha_1$. However, the exact value always is slightly larger than the approximated one: this discrepancy is due to the Joule heating, neglected in the approximate expression of Eq.~(\ref{Tm}), which raises a little the internal temperature of the structure. As the electrical current increases, the Joule heating cannot be neglected and the temperature of the junction is no longer controlled by the adjustment between convective and conductive heat fluxes. In such a case the reasoning made here is no longer valid and, as demonstrated in a recent article by Yang and coworkers \cite{Yang2013}, the effective figure of merit may overcome the value of the higher $ZT$ of the two materials considered for the device. However this situation where Joule heating is no longer negligible is very specific and it is not often encountered in classical designs of segmented thermoelectric generators, at least in constant parameters models such as the ones used in Ref.~\cite{Yang2013}. Indeed, in the generator regime, the electrical current is then limited by the value of the closed-circuit current since in such models the temperature difference $\Delta T = T_{\rm hot} - T_{\rm cold}$ is assumed to remain small in order to ensure linearity. The Joule heating thus seldom reaches sufficiently high values to have a significant influence on the system's behavior. This may explain why the conclusion by Bergman and Levy, i.e., that the global figure of merit of the system is always smaller than the figure of merit of each segment \cite{Bergman1991}, has been left unchallenged for more than 20 years.


\bibliographystyle{elsarticle-num}

\end{document}